\documentclass[prl,twocolumn,showpacs]{revtex4}

\usepackage{graphicx}
\usepackage{amsmath}

\newcommand{\e}{\mathrm{e}}
\newcommand{\ba}{{\mathbf{a}}}
\newcommand{\bat}{{\tilde{\mathbf{a}}}}

\begin{document}

\title
{Hierarchical Random Telegraph Signals in Nano-junctions with Coulomb Correlations}

\author
{P.-A. Bares and B. Braunecker}

\affiliation
{Institute of Theoretical Physics, Faculty of Basic Sciences,
Swiss Federal Institute of Technology Lausanne, EPFL, CH-1015
Lausanne }

%\date{22 February, 2002}
\date{\today}

\begin{abstract}
We propose a microscopic hamiltonian together with a master
equation description to model stochastic hierarchical Random
Telegraph Signal (RTS) or Pop-corn noise in nano-junctions. The
microscopic model incorporates the crucial Coulomb correlations
due to the trapped charges inside the junction or at the
metal-oxide interface. The exact solution of the microscopic model
is based on a generalization of the Nozi\`eres-De Dominicis method
devised to treat the problem of the edge singularity in the X-ray
absorption and emission spectra of metals. In the master equation
description, the experimentally accessible transition rates are
expressed in terms of the exact multi-channel Scattering matrix of
the microscopic hamiltonian.

\end{abstract}


\maketitle

In the last two decades, the experimental as well as the
theoretical investigations of noise in sub-micron structures have
developed into a new and fascinating subfield of mesoscopic
physics \cite{BlanterBuettiker00}. Recently a resurgence of
interest in tunneling through quantum dots and other
nano-structures has taken place, presumably due to progress in the
field of low-dimensional strongly correlated systems. As
electronic devices are reduced to sub-micron sizes, stochastic
fluctuations start to dominate their transport properties. 
In this letter we focus on 
generation-recombination noise that is often observed in
semi-conductors and is interpreted as due to fluctuations in the
number of carriers from the conduction (valence) band into traps
in the band gap. 
Random Telegraph Signal noise (RTS) or Pop-corn noise is a special
case of generation-recombination noise. The RTS may reflect the
transition of single ions or trapped electrons between various
meta-stable states inside the nano-structure. 

Tunneling through charge defects at interfaces or junctions has
been discussed in the literature earlier \cite{Schmidlin65,Caroli71}. In
particular, Matveev and Larkin showed that the current-voltage
characteristic of a tunnel-junction with a localized impurity
level exhibits a power-law singularity due to Coulomb interactions
between the charge carriers in the leads and the local charge
impurity \cite{MatveevLarkin92}. Remarkably enough, Geim et al.
and Cobden and Muzykantskii measured the Fermi-edge singularity in
resonant tunneling and in RTS noise \cite{Geim+94,Cobden+95}.
More recently Doudin et al. observed RTS fluctuations of up to
50\% of the average magneto-resistance in ultra-small junctions
\cite{Doudin+97}. Our investigations build further on these works.
We do not discuss here the problem of dephasing as referred to
in \cite{Zawadowski}.

\begin{figure}[p]
 \includegraphics{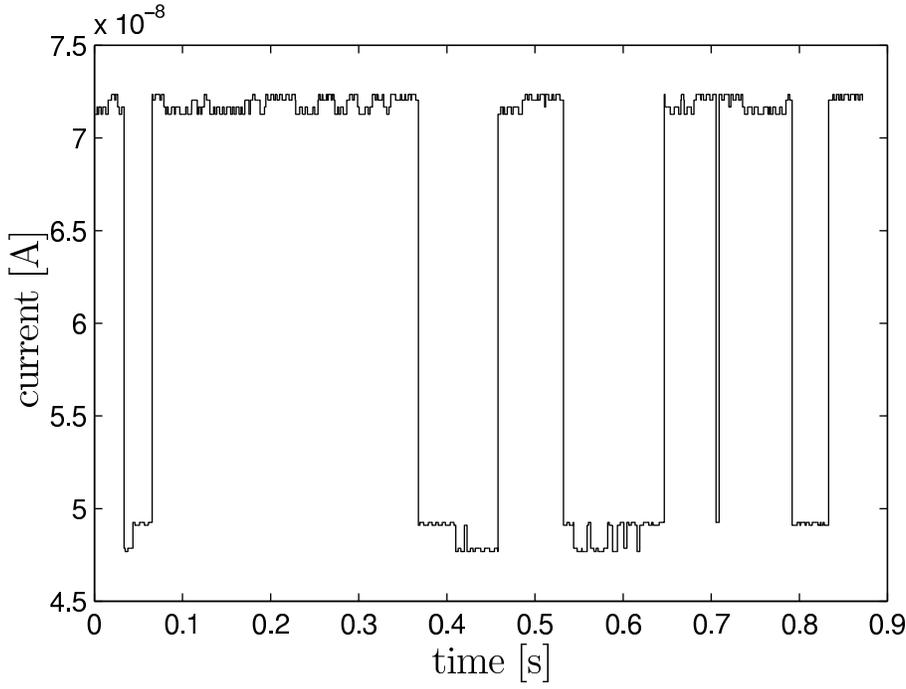}
 \caption{Computation of a three-level RTS current fluctuations using
 Eqs.~\eqref{lambda} and \eqref{current}.
 \label{fig1}}
\end{figure}

In this letter, we propose and solve a microscopic model as well
as a phenomenological description of Pop-corn noise observed in
tunneling through nano-junctions such as metal-oxide-metal
nano-structures (extensions of this work to
semi-conductor-oxide-semiconductor structures, etc. are possible).
At fixed chemical potentials, the current (conductance) jumps in a
random yet hierarchical manner between discrete levels (see Figure
\ref{fig1}). We argue that strong Coulomb correlations between trapped
charges located inside the junction (or at the metal-oxide
interface) and carriers in the leads govern the stochastic
switching between the different values of the current
(conductance). In the multi-channel and multi-trap case, we
express the transition rates in a master equation description in
terms of the parameters of the microscopic hamiltonian.

The model hamiltonian reads
\begin{eqnarray} 
H&=&\sum_{jj'}\epsilon^{jj'}
d_{j}^{\dagger}d_{j'}+\sum_{k \alpha}\epsilon_{k}^{\alpha}
c_{k\alpha}^{\dagger} c_{k\alpha}\cr &+&\sum_{jj'}U^{jj'}
d_{j}^{\dagger}d_{j}d_{j'}^{\dagger}d_{j'} 
\nonumber \\
&+&\sum_{k\dots\alpha'}
W_{kk'\alpha\alpha'}^{0}c_{k\alpha}^{\dagger} c_{k'\alpha'} 
\nonumber \\
&+&\sum_{k\dots\alpha'}
W_{kk'\alpha\alpha'}^{jj'}c_{k\alpha}^{\dagger}
c_{k'\alpha'}d_{j'}^{\dagger}d_{j}
\nonumber \\
&+&\sum_{kj\alpha}\left(V_{k\alpha}^{j}
c_{k\alpha}^{\dagger}d_{j}+h.c.\right)
\label{hamiltonian} 
\end{eqnarray}
where $c_{k\alpha}^{\dagger}$ creates an electron of momentum $k$
in lead $\alpha=L,R$; more generally the index $\alpha$ may also
label the various transverse channels $\alpha =1,\dots, n_c$
inside the leads. For simplicity's sake we shall omit here spin
degrees of freedom. $d_{j}^{\dagger}$ creates an electron in a
localized state $j$ inside the junction and at the interface.
Traps can fluctuate in space between different meta-stable
positions or migrate, yet we shall neglect these fluctuations.
$\epsilon^{jj'}$ denotes the energy of and transfer between
localized states. For simplicity, here we shall assume diagonal
$\epsilon^{jj'}=\delta_{jj'}E_j$. $\epsilon_{k}^{\alpha}$ denotes
the conduction band energies  and
$\mu_{\alpha}=\epsilon_{k_F}^{\alpha}$ the chemical potential in
channel $\alpha$. Moreover $\nu_{\alpha}$ is the density of states
at the $\alpha$ Fermi level. $U^{jj'}$ denotes the Coulomb
repulsion between localized states inside the junction. Because of
its crucial role in the hierarchical structure of tunneling, we
shall treat this term at the mean-field level
\cite{BrauneckerBares02}. $W_{kk'\alpha\alpha'}^{0}$ represents
the direct tunneling (non-diagonal matrix elements in the
$\alpha$ indices) and backscattering (diagonal matrix elements)
due to the metal-oxide interface. In the standard treatment of
tunneling, only the non-diagonal part of this term is taken into
account. The physical effect of the backscattering potential is a
renormalization of the momentum (energy)-dependent effective
density of states in the usual formula for the tunneling current.
$W_{kk'\alpha\alpha'}^{jj'}$ denotes the Coulomb scattering
between localized and extended states and represents the charge
assisted tunneling. The terms $W_{kk'\alpha\alpha'}^{0}$ and
$W_{kk'\alpha\alpha'}^{jj'}$ play a central role in this work and,
should we emphasize, shall be treated exactly below.
$V_{k\alpha}^{j}$ denotes the transfer integral between localized
states inside the junction (or at the interface) and extended
states in the leads.

To give a simple physical picture of the tunneling with trapping
we consider an hydrodynamical description of the electron fluids
in the leads. At wave-lengths that are much larger than the
average spacing between the particles, the charge and
magnetization density fluctuations are described by harmonic
oscillators that can be quantized. Coulomb correlations within the
leads in the charge and spin degrees of freedom are included in
this semi-classical heuristic argument. As charge $\delta Q$ and
magnetization $ \delta M$ are being trapped in the junction (or at
the interface), the electron fluids in the leads screen the
accumulated charge and magnetization. After trapping has taken
place, the energy increment consists in the charging and
magnetizing terms proportional to $\delta Q^2$, respectively,
$\delta M^2$, plus a Coulomb term between the charge and spin
fluctuations in the leads and the local charge and magnetization
in the junction (or at the interface): the charge and spin density
fluctuations can be described by displaced harmonic oscillators.
Hence the overlap between the de-trapped and the trapped
configurations is a Gaussian in the charge and magnetization
density fluctuations which vanishes in the infrared limit for
Coulomb potential and one-dimensional longitudinal modes. In
reality, the rate of charge $\delta\dot Q$ and magnetization
$\delta\dot M$ accumulation in the junction is proportional to the
charge and spin current density differentials between the
(channels) right and left leads. As a result
a dynamical cascade of Infrared Catastrophes occurs.

There are a number of experiments
\cite{Schmidlin65,Geim+94,Cobden+95,Doudin+97} where the Markov
property of the random switching process has been observed. Hence,
we consider a description in real time in terms of a master
equation. The idea is: i) to compute, for a fixed configuration
$\ba$ of trapped carriers, the charge assisted tunneling current
$\langle I\rangle_{\ba}$ at fixed chemical potentials
\cite{BrauneckerBares02}. The configuration vector
$\ba=(a_1,\dots,a_N)$ refers to $a_j=1$ if the local state
$j=1,\dots,N$ is occupied and $a_j=0$ otherwise; ii) to evaluate
the average current and the current-current correlation function
through the junction by weighting each configuration appropriately
\cite{BrauneckerBares02}. 
Subsequently we give expressions for the currents 
$\langle I\rangle_\ba$ and the current-current correlation
function.

Combining a projection technique with
partial tracing of bath (channels) degrees of freedom
\cite{Gardiner+00}, we arrive at a master equation for the
diagonal matrix elements of the local density operator, i.e., for
the probability density at time $t$ of being in a configuration
$\bf a$ as
\begin{equation} \label{master} 
\dot{p}({\ba},t)=-{\hat\lambda}^{\ba}p(\ba,t)
+\sum_{\bat\not=\ba}\lambda_{\ba}^{\bat}p(\bat,t)\;\;,\;\;
{\hat\lambda}^{\ba}=\sum_{\bat\not=\ba}\lambda^{\ba}_{\bat}
\end{equation}
with the transition rates from $\ba$ to $\bat$ given by
\begin{equation} \label{respfun}
	\lambda^{\ba}_{\bat} = 
	2 \mathrm{Re}\int_{0}^{\infty} dt {\rm Tr_B}
	\left[\rho_{B\ba}^{eq}\hat{H}_{\ba\bat}'(t)
	\hat{H}_{\bat\ba}'(0)\right]
\end{equation}
where $\rho_{B\ba}^{eq}$ represents the equilibrium density matrix
for the bath (channels) with trap configuration $\ba$. Notice that
$\ba$ and $\bat$ differ by a single site occupation. Further
$\hat{H}_{\ba\bf b}'=P_{\ba}\hat{H}'P_{\bf b}$ denotes a projected
interaction picture of the transfer term $H'=\sum_{k\alpha
}\left(V_{k\alpha}^{j} c_{k\alpha}^{\dagger}d_{j}+ h.c.\right)$.
The Markov approximation underlying equation \eqref{master} requires the
correlation time of the response function in equation \eqref{respfun} at
finite temperature and in the presence of weak disorder to be much
shorter than the characteristic time variations of the junction
density matrix. For an arbitrary number of localized states the
master equation \eqref{master} can be solved, in principle, by a method due
to Kirchhoff \cite{Kirchhoff1844}. A systematic Graph theoretic
technique has also been developed by Weidlich \cite{Weidlich78}.
Figure \ref{fig1} illustrates a three level hierarchical RTS noise. The
crucial point here is that the transition rates, which are
experimentally accessible, can be expressed in terms of the local
correlation function of the trapped states in the junction (or at
the interface).

At this stage we have to tackle Dyson's equation for the transient
propagator and from it infer the response function \eqref{respfun}. To solve
Dyson's equation a substantial generalization of a method
developed by Nozi\`eres and De Dominicis in the context of the
X-ray edge problem in metals \cite{NozieresDeDominicis69,x-ray} is
required. In the present case, Dyson's equation reduces to a
system of singular integral equations that can be solved exactly
by the techniques of Muskhelishvili and Vekua
\cite{Muskhelishvili53,Vekua67,BrauneckerBares02} (see also
\footnote{The same equation arises in the context of 
the non-equilibrium X-ray problem for which 
Ng \cite{Ng} has proposed a solution. 
However, Combescot and Roulet \cite{CombescotRoulet} pointed
out that there were some inconsistencies in Ng's results without
being able to resolve them. In our opinion, the difficulties result
from an incorrect ansatz to the solution of the Dyson equation.
They do not appear in the present treatment.}).
While the mathematics is rather involved let us attempt to explain
in simple terms the essence of the method. We consider the process
of single electron transfer into and off the trap state $j_0$ by
$H'$, the occupation of all other localized states being fixed.
The potential $W_{\ba}$ acting on the conduction electrons can be
written as the sum of all $W^j_{kk'}$ (considered as matrix in the
indices $\alpha\alpha'$) for which $a_{j_0}=1$ plus the direct
tunneling and backscattering term $W^0_{kk'}$. A single electron
transfer leads to $W_{\bat}$, where ${\bat}$ differs from ${\ba}$
by the occupation of the single site $j_0$. Immediately after this
single electron transfer the new potential felt by the conduction
electrons reads $\left[W^{\ba}_{\bat}\right]_{kk'}=\left[ W_{\bat
}\right]_{kk'}-\left[W_{\ba}\right]_{kk'}$. So the task can be
decomposed into two steps, namely compute the propagator for a
particle subject to the potential $W_{\ba}$, and then extract the
transient propagator for the potential $W^{\ba}_{\bat}$ which acts
during a finite time interval. This transient propagator allows to
obtain the response function required to compute the transition
rate \eqref{respfun}. Following Nozi\`eres and De Dominicis, we assume
separable potentials $W_{kk'\alpha\alpha'}^{0}=W_0^{\alpha\alpha'}
u_{k\alpha}u_{k'\alpha'}$,
$W_{kk'\alpha\alpha'}^{jj'}=\delta_{jj'} W^{\alpha\alpha'}_j
u_{k\alpha}u_{k'\alpha'}$ , and $V_{k\alpha}^j=V_{\alpha}^j
u_{k\alpha} $ where $u_{k\alpha}$ is a cut-off function centered
around the $\alpha$ Fermi surface, and take into account S-wave
scattering only. This factorization simplifies the mathematics
without affecting the physics, as it allows to perform readily the
sums over $k$ to obtain the local propagators at the junction. With
these assumptions, the finite temperature transition rate for an
arbitrary number of traps, including direct tunneling, reads
\begin{align} 
    \lambda^\ba_\bat &= 2 \pi \ \mathrm{Re} \sum_{\alpha_1 \kappa}
    \left(\frac{i\xi_0\beta}{2 \pi}\right)^{\epsilon_{\alpha_1}}
    \nonumber \\
    &\times
    \frac{2 \Gamma(\epsilon_{\alpha_1})
    \mathcal{A}_{\alpha_1 \kappa} }%
         {\left|\Gamma\left(\frac{\epsilon_{\alpha_1}+1}{2} +
         i \frac{\beta}{2 \pi} | \widetilde{E}^\ba_\bat \mp \mu_{\kappa}|\right)
         \right|^2} \nonumber \\
    &\times
        \frac{\cosh\left(\frac{\beta}{2}(\widetilde E^\ba_\bat \mp
        \mu_{\kappa})-
              i \frac{\pi}{2} \epsilon_{\alpha_1}\right)}%
             {\cosh(\beta(\widetilde{E}^\ba_\bat \mp\mu_{\kappa})) +
             \cos(\pi \epsilon_{\alpha_1})}
	\label{lambda}			 
\end{align}
where $\xi_0$ is of the order of the conduction bandwidths,
$\Gamma(x)$ denotes Euler's Gamma function and $\beta = 1/k_B T$
the inverse temperature. The $-$ sign refers to trapping
(absorption) and the $+$ sign to de-trapping (emission).
$\kappa = 1, \dots, m_c (\le n_c)$ labels distinct chemical potentials
$\mu_\kappa$.
$\widetilde{E}^\ba_\bat={\widetilde E}_{\bat}-{\widetilde
E}_{\ba}$ with ${\widetilde E}_{\ba}$ being the energy of the
localized state renormalized (closed loop contributions) by the
potential $W_{\ba}$, i.e., ${\widetilde E}_{\ba}=E_{\ba}+\Delta
E_{\ba}^0=\sum_{j=1}^{N}a_j E_j+\Delta E_{\ba}^0$. The
local Coulomb repulsion between trapped charges can be
incorporated in the mean field approximation and contributes an
additional term $U N_{\ba}(N_{\ba}-1)/2$ to the energy
${\widetilde E}_{\ba }$, with $N_{\ba}$ the number of trapped
charges in configuration $\bf a$, and $U$ the average potential in
the junction. Furthermore $\mathcal{A}_{\alpha_1 \kappa}$ is given by
\begin{multline}
    \mathcal{A}_{\alpha_1 \kappa} = -i \sum_{\alpha \alpha'} V^{j_0}_\alpha
    (D_{\bat+}^{-1} D_{\ba+} (B^\ba_\bat)^{-1})_{\alpha \alpha_1} \\
	(B^\ba_\bat D_{\ba+}^{-1} D_{\bat+} \Lambda^\bat_\kappa)_{\alpha_1 \alpha'}
    V^{j_0}_{\alpha'}
\end{multline}
where $j_0$ denotes the trap subject to the change in occupation
number $(a_{j_0} \neq \tilde{a}_{j_0})$. $\nu_{\alpha \alpha'} =
\nu_\alpha \delta_{\alpha \alpha'}$ is the density of states at
the Fermi level $\mu_\alpha$ and we have defined the matrices
(assuming summation over $\hat{\alpha}$)
\begin{align}
    (D_{\ba\pm})_{\alpha \alpha'} &= 
	[\pi (\tan\theta_\alpha \pm i) \nu_\alpha]^{-1} \delta_{\alpha \alpha'}
	+ W_\ba^{\alpha \alpha'}\\
    \Lambda^\bat_{\kappa} &= [\Sigma^\bat_{\kappa}]^{-1}  \chi_\kappa 
	                           \nu W_\bat [\Sigma^\bat_{\kappa-1}]^{-1} (\nu W_\bat)^{-1}\\
    (\Sigma^\bat_\kappa)_{\alpha \alpha'} &= \delta_{\alpha \alpha'} +
	\pi (\tan\theta_{\alpha} + i \sigma(\alpha,\kappa) )
    \nu_{\alpha} W_\bat^{\alpha \alpha'}
\end{align}
with $\sigma(\alpha,\kappa) = +1$ if $\mu_\alpha <
\mu_{\kappa}$ and $-1$ if $\mu_\alpha \ge
\mu_{\kappa}$ and 
$(\chi_\kappa)_{\alpha \alpha'} = 
\delta_{\alpha \alpha'} \delta_{\mu_\alpha, \mu_\kappa}$. 
We furthermore have assumed that 
$\mu_1 \ge \mu_2 \ge \dots \ge \mu_{n_c}$. 
The angle $\theta_\alpha$ parameterizes the
short time behavior of the single particle propagator in channel
$\alpha$ \cite{NozieresDeDominicis69}. The matrix
$\mathcal{S^\ba_\bat} = 
D_{\ba+}^{-1} D_{\bat+} D_{\bat-}^{-1} D_{\ba-}$ is related to
the generalized multi-channel Scattering matrices, 
$\mathcal{S}_\ba = D_{\ba+} D_{\ba-}^{-1}$, and is
diagonalized by $B^\ba_\bat$, $\mathcal{S^\ba_\bat}= B^\ba_\bat
\e^{2i \delta^\ba_\bat} (B^\ba_\bat)^{-1}$, with the real
eigenvalues $\delta^\ba_\bat =
\mathrm{diag}((\delta^\ba_\bat)_1,\dots,(\delta^\ba_\bat)_{n_c})$.
We use the notation $\epsilon_{\alpha_1} = 
\pm 2 (\delta^\ba_\bat)_{\alpha_1}/\pi -
\mathrm{Tr}((\delta^\ba_\bat)^2)/\pi^2$.

In the low temperature limit $k_B T\ll
|\widetilde{E}_{\bat}^{\ba}\mp\mu_{\alpha_1}|$
the transition rates show the power-law divergence
with the exponents $\epsilon_{\alpha_1}$
known from the X-ray problem \cite{NozieresDeDominicis69,x-ray}.
The first correction in $T$ is proportional to
$(k_B T / | \widetilde{E}^\ba_\bat \mp \mu_{\kappa}|)^2$.
In the high temperature limit, 
$k_B T \gg |\widetilde{E}_{\bat}^{\ba}\mp\mu_{\alpha_1}|$,
$\lambda$ shows a power-law behavior in $T$,
\begin{equation}
	\lambda^\ba_\bat \sim \sum T^{-\epsilon_{\alpha_1}}.
\end{equation}

\begin{figure}[p]
 \includegraphics{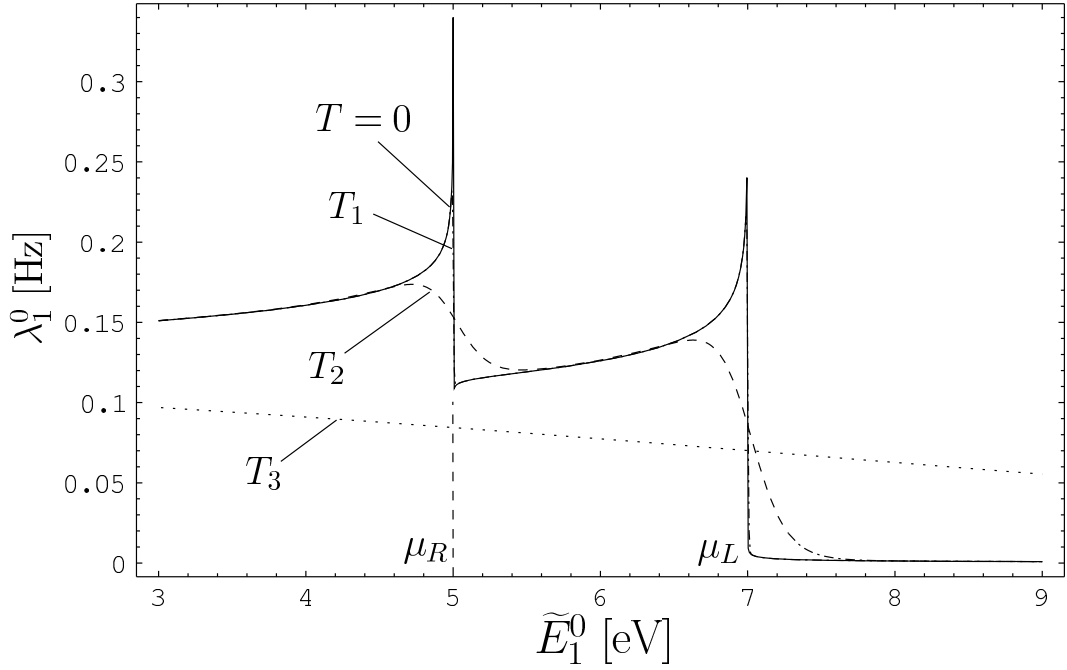}
 \caption{Typical transition rate behavior in function of the energy 
 of the localized state for the case of a single trap (Eq.~\eqref{lambda}).
 The different curves correspond to the temperatures
 $T=0, T_1 = T_3^{1/3} \approx 40 \, \mathrm{K}, T_2 = T_3^{2/3} \approx 1500 \, \mathrm{K},
 T_3 = \mu_R / k_B \approx \mbox{60,000} \, \mathrm{K}$.
 \label{fig2}}
\end{figure}

Specializing to a single trap, a single channel, and 
in the absence of direct tunneling the finite temperature
expression \eqref{lambda} reduces to the result
of \cite{Cobden+95} (their Eq. (5)).
The effect of finite temperature is a broadening of the
edge singularity, as illustrated in Figure \ref{fig2}. 
A similar broadening occurs in the case of weak disorder 
in weakly coupled electrodes: The exponential decay
of the electron propagators in the electrodes leads
to an imaginary contribution to the singular
denominator, 
$| \widetilde{E}^\ba_\bat \pm \mu_\kappa + i \gamma_\alpha |$,
where $\gamma_\alpha = \pi \nu_\alpha n_{\text{dis},\alpha} U^2_{\text{dis},\alpha}$
and $n_{\text{dis},\alpha}, U_{\text{dis},\alpha}$ are the impurity density
and the weak disorder potential in the channel $\alpha$, respectively.

For a fixed configuration $\ba$ it is possible to adapt the method
provided by Caroli et al. \cite{Caroli71} to the calculation
of the tunneling current in the presence of the backscattering 
$W_\ba^{\alpha \alpha'} (\alpha \neq \alpha')$.
The result for the simplest case of two channels $\alpha = L,R$ 
is given by
\begin{equation} \label{current}
	\langle I\rangle_\ba = e  \int \frac{d\omega}{2 \pi}
	\frac{W_\ba^{LR}\rho_L^h(\omega) \rho_R^e(\omega) 
	    - W_\ba^{RL} \rho_R^h(\omega) \rho_L^e(\omega)}%
	{\mathcal{D}^r(\omega) \mathcal{D}^a(\omega)},
\end{equation}
where $e$ is the electron charge, 
$\rho_\alpha^h(\omega), \rho_\alpha^e(\omega)$ represent 
the density of holes and electrons in the electrode $\alpha$
respectively, and
\begin{multline}
	\mathcal{D}^{r,a}(\omega) = 
	(1-W_\ba^{LL} g_L^{r,a}(\omega))
	(1-W_\ba^{RR} g_R^{r,a}(\omega)) \\ 
	- W_\ba^{LR} W_\ba^{RL} g_L^{r,a}(\omega) g_R^{r,a}(\omega),
\end{multline}
with $g_\alpha^{r,a}$ the retarded and advanced free Green's
functions in channel $\alpha$.
As an illustration for the current-current correlation function,
$S_{II}(t)$, we choose a system with a single trap, $\ba = a = 0,1$.
In the master equation approach $S_{II}$ becomes
\begin{equation}
	S_{II}(t) = \e^{-(\lambda^0_1 + \lambda^1_0) |t|} 
	\frac{\lambda^0_1 \lambda^1_0}{(\lambda^0_1 + \lambda^1_0)^2}
	(\langle I \rangle_1 - \langle I \rangle_0)^2.
\end{equation}

In summary, we have been able to evaluate, for an arbitrary number of
trapped charges, the transition rates of the master equation
description of RTS noise and the tunneling current 
in terms of microscopic model hamiltonian parameters. 
We emphasize that this approach is not restricted to two-level systems 
which is often found in the literature \cite{Zawadowski}, but allows
for several fluctuators that can create a more complex, in particular
an hierarchical, RTS noise.
In future work, we intend to take into account the
correlations in the leads (channels) as well as spin degrees of
freedom. The latter are required, for example, to model RTS noise
in magnetic nano-wires due to charge and magnetization trapping
inside the junction.

%
%%%%%%%%%%%%%%%%%%%%%%%%%%%%%%%%%%%%%%%%%%%%%%%%%%%%%%%%%%%%%%%%%%%%
%
%
\section*{ACKNOWLEDGMENTS}
We thank B. Altshuler, J.-Ph. Ansermet, P.A. Lee, Ph. Nozi\`eres,
T.M. Rice, A. Tsvelik and X.-G. Wen for discussions. The support
of the Swiss National Fonds is gratefully acknowledged.


\end{document}